\documentclass[lettersize,journal]{IEEEtran}
\usepackage{amsmath,amsfonts}
\usepackage{algorithmic}
\usepackage{algorithm}
\usepackage{float} 
\usepackage{array}
\usepackage{textcomp}
\usepackage{stfloats}
\usepackage{url}
\usepackage{verbatim}
\usepackage{graphicx}
\usepackage{subcaption}
\usepackage{cite}
\usepackage{amsmath}
\usepackage{xcolor}
\usepackage{amsmath}
\usepackage{amssymb}
\usepackage{graphicx}
\begin{document}

\title{QoS-Constrained Sum-Rate Maximization in RIS-Assisted Pinching-Antenna NOMA Systems}

\author{Lütfullah Özkan, Saliha Büyükçorak
\vspace{-6mm}
\thanks{
L. Ozkan and S. Buyukcorak are with the Department of Electronics Engineering, Gebze Technical University, Kocaeli, Turkey (e-mail: \{l.ozkan, sbuyukcorak\}@gtu.edu.tr).
}

}

\markboth{IEEE WIRELESS COMMUNICATIONS LETTERS, VOL. XXX, XXX}%
{Shell \MakeLowercase{\textit{et al.}}: A Sample Article Using IEEEtran.cls for IEEE Journals}


\maketitle

\begin{abstract}
This paper investigates a two-user downlink non-orthogonal multiple access (NOMA) system jointly assisted by a single-waveguide pinching-antenna system (PASS) and a reconfigurable intelligent surface (RIS). To maximize the achievable sum rate under quality-of-service (QoS) constraints, the pinching antenna (PA) positions, RIS phase shifts, and power allocation coefficients are jointly optimized. The resulting non-convex optimization problem is addressed using an alternating optimization (AO) framework, where power allocation is obtained in closed form, while RIS phase shifts and PA positions are optimized through majorization-minimization (MM) and sequential quadratic programming (SQP), respectively. Numerical results show that the proposed architecture achieves higher sum rates than conventional-antenna and PASS-only benchmarks. Moreover, the MM- and SQP-based optimization stages achieve performance comparable to semidefinite relaxation (SDR) and grid-based exhaustive search, respectively, while the AO algorithm exhibits rapid convergence.
\end{abstract}


\vspace{-0.6cm}
\section{Introduction}

Flexible-antenna systems have attracted considerable attention in wireless communications due to their ability to adapt antenna configurations to propagation conditions~\cite{Ding2025PASSPerspective}. Among these technologies, pinching-antenna systems (PASS) enable antenna positions to be adjusted along a dielectric waveguide. A PASS consists of a waveguide and pinching antennas (PAs), formed by placing small dielectric elements at selected positions along the waveguide~\cite{Liu2025PASSArchitecture}. Unlike conventional fixed-antenna (FA) systems, PASS provides additional flexibility in shaping wireless channels by adjusting the radiation points~\cite{Liu2026PASSTutorial}. Recent studies have investigated waveguide structures and PA activation methods~\cite{Xu2026GeneralizedPASS}, as well as PA positioning for single-user downlink transmission considering path loss and signal phase~\cite{Xu2025DownlinkPASS}. For multi-user transmission, power-domain non-orthogonal multiple access (NOMA) can be employed to serve users over shared time-frequency resources~\cite{Ding2025PASSPerspective}. In NOMA, user signals are superimposed with different power allocation coefficients, while successive interference cancellation (SIC) is employed to mitigate inter-user interference during decoding. Accordingly, joint PA positioning and power allocation have been investigated in PA-NOMA systems~\cite{Zhou2025NOMAPASS}. Related studies have also addressed matching-based PA activation~\cite{Wang2025PASSActivation}, QoS-constrained two-user transmission~\cite{Xu2025QoSNOMA}, and resource allocation and precoding in multi-waveguide systems~\cite{Wang2025MultiWaveguideActivation,Hu2025MultiWaveguideNOMA}. 

Reconfigurable intelligent surface (RIS) offers another means of controlling wireless propagation by adjusting the phase shifts of reflected signals. While PASS enables the adjustment of radiation points, RIS controls the phases of reflected signals, making the two technologies complementary. Their integration has been investigated for single-user downlink transmission through the joint optimization of PA positions and RIS phase shifts~\cite{Fu2026RISPASS}. However, extending this architecture to multi-user NOMA transmission requires coordinating PA positioning and RIS phase control with power allocation under user-specific QoS requirements. To the best of our knowledge, the joint optimization of these three variable sets for sum-rate maximization in a two-user RIS-assisted PASS-NOMA system under QoS constraints has not yet been investigated. Motivated by this research gap, the main contributions of this paper are summarized as follows:
\begin{itemize}
\item We formulate a QoS-constrained sum-rate maximization problem for a two-user RIS-assisted PASS-NOMA system, jointly optimizing PA positions, RIS phase shifts, and power allocation coefficients. 
\item To address the resulting non-convex problem, we develop an alternating optimization (AO) framework that decomposes it into three subproblems. A closed-form solution is derived for power allocation, while majorization-minimization (MM) and sequential quadratic programming (SQP) are employed for RIS phase shifts and PA positions, respectively.
\item We evaluate the  sum rate and convergence behavior of the proposed algorithm through numerical simulations. The MM- and SQP-based optimization stages achieve performance comparable to semidefinite relaxation (SDR) and grid-based exhaustive search, respectively. Comparisons with conventional-antenna and PASS-only benchmarks demonstrate the sum-rate gains achieved by jointly employing PASS and RIS in the NOMA system.
\end{itemize}

\section{System Model}

\begin{figure}[!t]
	\centering
	\includegraphics[width=3.5in]{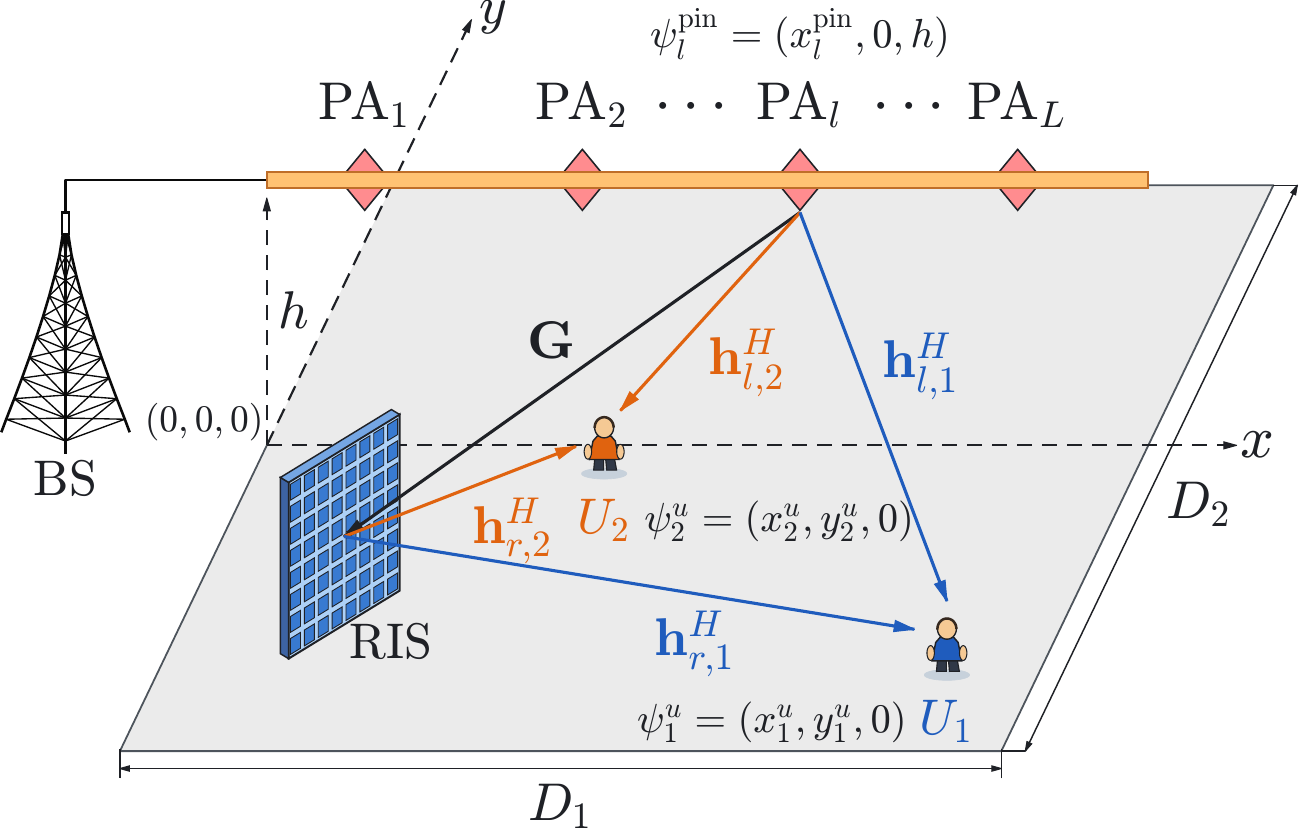}
    \vspace{-2mm}
	\caption{\textcolor{black}{An RIS-assisted downlink PA-NOMA system model.}}
	\label{sys_mod}
    \vspace{-8mm}
\end{figure} 

This paper considers an RIS-assisted PA system employing power-domain downlink NOMA, as illustrated in Fig. \ref{sys_mod}. A base station (BS) serves two single-antenna users through $L$ PAs deployed along a single dielectric waveguide, with the assistance of a passive RIS.
The waveguide extends along the \(x\)-axis over a length of \(D_1\) at a height \(h\). The position of the \(l\)-th PA is given by \(\boldsymbol{\psi}_l^{\mathrm{pin}}=(x_l^{\mathrm{pin}},0,h)\), where \(x_l^{\mathrm{pin}}\in[0,D_1]\) and \(l\in\mathcal{L}=\{1,2,\ldots,L\}\), and the PA positions are denoted by \(\mathbf{x}^{\mathrm{pin}}=[x_1^{\mathrm{pin}},\ldots,x_L^{\mathrm{pin}}]^T\). 
The RIS is modeled as a uniform planar array (UPA) consisting of $M=M_r\times M_c$ passive reflecting elements, where $M_r$ and $M_c$ denote the numbers of rows and columns, respectively. The RIS phase shift matrix is defined as $\boldsymbol{\Phi}=
\mathit{diag}(e^{{j}\theta_1},
\ldots,e^{{j}\theta_m},\ldots,e^{{j}\theta_M})$, where $\theta_m\in[0,2\pi)$ denotes the phase shift introduced by the $m$-th RIS element. With the RIS reference element located at \(\boldsymbol{\psi}^{\mathrm{ris}}=(x^{\mathrm{ris}},y^{\mathrm{ris}},h)\), the position of the \(m\)-th RIS element is given by \(\boldsymbol{\psi}_m^{\mathrm{ris}}=(x^{\mathrm{ris}},y^{\mathrm{ris}}+(m-1-(i-1)M_r)d,h-(i-1)d)\), for \(m\in\mathcal{M}=\{1,2,\ldots,M\}\), where \(i=\lfloor \frac{m-1}{M_r}\rfloor+1\) and \(d\) denotes the spacing between adjacent RIS elements.
The two users are located within a rectangular area of size \(D_1\times D_2\), with positions \(\boldsymbol{\psi}_n^u=(x_n^u,y_n^u,0)\), where \(x_n^u\in[0,D_1]\), \(y_n^u\in[0,D_2]\), and \(n\in\mathcal{N}=\{1,2\}\). 
Let $d_{ln}^{\mathrm{pu}}= \|\boldsymbol{\psi}_n^u-\boldsymbol{\psi}_l^{\mathrm{pin}}\|$,
$d_{lm}^{\mathrm{pr}}=
\|\boldsymbol{\psi}_m^{\mathrm{ris}}-
\boldsymbol{\psi}_l^{\mathrm{pin}}\|$, and
$d_{mn}^{\mathrm{ru}}=
\|\boldsymbol{\psi}_n^u-
\boldsymbol{\psi}_m^{\mathrm{ris}}\|$ denote the distances from the $l$-th PA to user $U_n$, from the $l$-th PA to the $m$-th RIS element, and from the $m$-th RIS element to $U_n$, respectively.
Following the spherical-wave channel model in \cite{Fu2026RISPASS},
the direct channel vector between the PAs and user $U_n$,
$\mathbf h_{d,n}\in\mathbb C^{L\times1}$, is expressed as
\begin{equation}
\resizebox{0.9\columnwidth}{!}{$\displaystyle
\mathbf h_{d,n}=
\left[
\frac{\sqrt{\eta}e^{-j\kappa_0d_{1n}^{\mathrm{pu}}}}
{d_{1n}^{\mathrm{pu}}},
\ldots,
\frac{\sqrt{\eta}e^{-j\kappa_0d_{ln}^{\mathrm{pu}}}}
{d_{ln}^{\mathrm{pu}}},
\ldots,
\frac{\sqrt{\eta}e^{-j\kappa_0d_{Ln}^{\mathrm{pu}}}}
{d_{Ln}^{\mathrm{pu}}}
\right]^T,$}
\end{equation}
where $\eta=\left(\frac{c}{4\pi f_c}\right)^2$,
$\kappa_0=\frac{2\pi f_c}{c}$, $c$ is the speed of light,
and $f_c$ is the carrier frequency. The PA-RIS channel matrix is denoted by $\mathbf{G}\in\mathbb{C}^{M\times L}$,
whose $(m,l)$-th element is given by
\begin{equation}
[\mathbf{G}]_{m,l}
=
\frac{\sqrt{\eta}e^{-j\kappa_0d_{lm}^{\mathrm{pr}}}}
{d_{lm}^{\mathrm{pr}}}.
\end{equation}
Similarly, the RIS-user channel vector is
denoted by $\mathbf{h}_{r,n}\in\mathbb{C}^{M\times1}$,
whose $m$-th element is given by
\begin{equation}
[\mathbf{h}_{r,n}]_m
=
\frac{e^{-j\kappa_0d_{mn}^{\mathrm{ru}}}}
{d_{mn}^{\mathrm{ru}}}.
\end{equation}
With power-domain NOMA, the BS superimposes the signals intended for the two users before feeding them into the common waveguide. The resulting signal is expressed as $s=\sum_{n=1}^{2}\sqrt{\alpha_n}s_n$, where $\alpha_n$ is the power allocation coefficient of $U_n$, and $s_n$ is the information symbol satisfying $\mathbb{E}[|s_n|^2]=1$. The power allocation coefficients satisfy $\alpha_1>\alpha_2>0$ and $\sum_{n=1}^{2}\alpha_n=1$. 
As the superimposed signal propagates along the waveguide, it experiences position-dependent phase shifts before being radiated by the PAs. For analytical tractability, the in-waveguide propagation loss is neglected, while the propagation-induced phase shifts are retained \cite{Xu2026GeneralizedPASS}. Accordingly, assuming ideal equal-power distribution among the PAs, the radiated signal vector is expressed as
\begin{equation}
\mathbf{w}=\sqrt{\frac{P}{L}}
\left[
e^{-{j}\varphi_1},
\ldots,
e^{-{j}\varphi_l},
\ldots,
e^{-{j}\varphi_L}
\right]^T s,
\label{eq:waveguide}
\end{equation}
where $P$ is the total radiated power and $\varphi_l=\frac{2\pi}{\lambda_g}\|\boldsymbol{\psi}_0^{\mathrm{pin}}-
\boldsymbol{\psi}_l^{\mathrm{pin}}\|$ is the phase shift caused by signal propagation along the waveguide. Here, $\boldsymbol{\psi}_0^{\mathrm{pin}}$ denotes the waveguide feed point, $\lambda_g=\frac{\lambda}{n_{\mathrm{eff}}}$ is the guided wavelength, and $n_{\mathrm{eff}}$ is the effective refractive index of the dielectric waveguide.

The signal received by $U_n$ is expressed as
\begin{equation}
y_n=
\left(
\mathbf{h}_{r,n}^H\boldsymbol{\Phi}\mathbf{G}
+\mathbf{h}_{d,n}^H
\right)\mathbf{w}+z_n,
\label{eq:received_signal}
\end{equation}
where $z_n\sim\mathcal{CN}(0,\sigma^2)$ denotes the additive white Gaussian noise. Substituting $\mathbf{w}$ from \eqref{eq:waveguide} into \eqref{eq:received_signal}, the received signal can be rewritten as
\begin{equation}
y_n=\sqrt{\frac{P}{L}}\,g_n
\sum_{k=1}^{2}\sqrt{\alpha_k}s_k+z_n,
\label{eq:received_signal_noma}
\end{equation}
where $g_n$ is the effective channel coefficient of $U_n$, given by
\begin{equation}
\resizebox{0.85\columnwidth}{!}{$\displaystyle
\begin{aligned}
g_n
&\triangleq
\sum_{l=1}^{L}
\frac{\sqrt{\eta}\,e^{{j}(\kappa_0d_{ln}^{\mathrm{pu}}-\varphi_l)}}
{d_{ln}^{\mathrm{pu}}}\\[2mm]
&\quad+
\sum_{m=1}^{M}
\frac{e^{{j}(\theta_m+\kappa_0d_{mn}^{\mathrm{ru}})}}
{d_{mn}^{\mathrm{ru}}}
\sum_{l=1}^{L}
\frac{\sqrt{\eta}\,e^{-{j}(\kappa_0d_{lm}^{\mathrm{pr}}+\varphi_l)}}
{d_{lm}^{\mathrm{pr}}}.
\end{aligned}
$}
\label{eq:effective}
\end{equation}
For subsequent optimization, the effective channel in \eqref{eq:effective} can be equivalently written as
\begin{equation}
g_n=\mathbf{v}^T\mathbf{a}_n+b_n,
\label{eq:effective_compact}
\end{equation}
where
$\mathbf{v}=
[e^{j\theta_1},\ldots,e^{j\theta_m},\ldots,
e^{j\theta_M}]^T$,
$\mathbf{a}_n=
\mathit{diag}(\mathbf{h}_{r,n}^H)
\mathbf{G}\mathbf{u}$,
$b_n=\mathbf{h}_{d,n}^H\mathbf{u}$,
and
$\mathbf{u}=
[e^{-j\varphi_1},\ldots,e^{-j\varphi_l},\ldots,
e^{-j\varphi_L}]^T$.

Unlike conventional FA systems, the PA positions and RIS phase shifts can be adjusted to improve the users' channel conditions. For a single user, placing the PAs closer to the user can reduce path loss and improve the achievable rate \cite{Fu2026RISPASS}. However, in the two-user NOMA system, PA positions and RIS phase shifts must be jointly optimized to maximize the sum rate while satisfying the users' QoS requirements.
The users are ordered according to their effective channel gains, such that $|g_2|^2\geq|g_1|^2$, where $U_1$ and $U_2$ denote the weak and strong users, respectively. For notational convenience, we define
$\gamma_n=P|g_n|^2/(L\sigma^2)$ as the channel SNR parameter
of $U_n$. Accordingly, $U_1$ decodes its own signal while treating $U_2$'s signal as interference. Its achievable rate is given by
\begin{equation}
R_1=\log_2\left(
1+\frac{\alpha_1\gamma_1}
{\alpha_2\gamma_1+1}
\right).
\label{eq:R1}
\end{equation}
At $U_2$, SIC is performed by first decoding $U_1$'s signal. The corresponding achievable rate is $R_{2\rightarrow1}=\log_2\left(
1+\frac{\alpha_1\gamma_2}
{\alpha_2\gamma_2+1}
\right).$ Under the assumed channel ordering, $\gamma_2\geq\gamma_1$, we have $R_{2\rightarrow1}\geq R_1$, ensuring that $U_2$ can decode $U_1$'s signal at its achievable rate. After canceling $U_1$'s signal under perfect SIC, $U_2$ decodes its own signal at the achievable rate
\begin{equation}
R_2=\log_2(1+\alpha_2\gamma_2).
\label{eq:R2}
\end{equation}
To satisfy the QoS requirements of both users, $R_1\geq R_1^{\min}$ and $R_2\geq R_2^{\min}$ must hold, where $R_1^{\min}$ and $R_2^{\min}$ denote the minimum required rates of $U_1$ and $U_2$, respectively.

\vspace{-0.2cm}
\section{Problem Formulation and Proposed Solution}

The objective is to maximize the achievable sum rate by jointly optimizing the power allocation coefficients, PA positions, and RIS phase shifts while satisfying the QoS requirements of both users. Using \eqref{eq:R1} and \eqref{eq:R2}, the sum rate is expressed as
\begin{equation}
R_{\mathrm{sum}}=R_1+R_2=
\log_2\left(
\frac{(1+\gamma_1)(1+\alpha_2\gamma_2)}
{1+\alpha_2\gamma_1}
\right).
\label{eq:sumrate}
\end{equation}
For notational convenience, we define
\begin{equation}
\mathcal F(\alpha_2,\mathbf x^{\mathrm{pin}},\mathbf v)
\triangleq
\frac{(1+\gamma_1)(1+\alpha_2\gamma_2)}
{1+\alpha_2\gamma_1}-1.
\label{eq:F}
\end{equation}
such that
$R_{\mathrm{sum}}=\log_2(1+\mathcal F)$. Since the logarithm is strictly increasing, maximizing the sum rate is equivalent to
maximizing $\mathcal F$. The optimization problem
is formulated as
\begin{subequations}
\label{eq:P1}
\begin{align}
\text{\textbf{(P1)}}\quad
&\max_{\alpha_2,\mathbf{x}^{\mathrm{pin}},\mathbf v}
\mathcal F(\alpha_2,\mathbf{x}^{\mathrm{pin}},\mathbf v)
\label{eq:P1a}\\
\text{\textbf{s.t.}}\quad
&C_1:\ x_l^{\mathrm{pin}}
-x_{l-1}^{\mathrm{pin}}\geq\Delta,
\quad \forall l\in\mathcal L\setminus\{1\},
\label{eq:P1b}\\
&C_2:\ 0\leq x_l^{\mathrm{pin}}
\leq D_1,\quad\forall l\in\mathcal L,
\label{eq:P1c}\\
&C_3:\ |v_m|=1,\quad\forall m\in\mathcal M,
\label{eq:P1d}\\
&C_4:\ 0<\alpha_2<\frac12,
\label{eq:P1e}\\
&C_5:\ R_1\geq R_1^{\min},
\label{eq:P1f}\\
&C_6:\ R_2\geq R_2^{\min},
\label{eq:P1g}\\
&C_7:\ |g_2|^2\geq|g_1|^2.
\label{eq:P1h}
\end{align}
\end{subequations}
Here, $\alpha_1=1-\alpha_2$. Constraint $C_1$ imposes a minimum spacing $\Delta$ between adjacent PAs to mitigate mutual coupling, while $C_2$ restricts their positions to the waveguide. $C_3$ enforces the unit-modulus condition of the RIS elements, and $C_4$ specifies the power allocation range consistent with $\alpha_1>\alpha_2>0$. $C_5$ and $C_6$ enforce the minimum rate requirements of the two users, respectively, while $C_7$ maintains the channel ordering underlying the SIC decoding order.
Problem \textbf{P1} is non-convex due to the nonlinear dependence of the effective channels on the PA positions and RIS phase shifts, the coupling among the variables, and the unit-modulus constraints. To address this problem, we adopt an AO approach, where the power allocation coefficients, RIS phase shifts, and PA positions are optimized sequentially with the other variables fixed.

\vspace{-0.4cm}
\subsection{Power Allocation Optimization}

For fixed PA positions and RIS phase shifts, the power allocation problem reduces to the optimization of $\alpha_2$, since $\alpha_1=1-\alpha_2$. Let $\beta_n=2^{R_n^{\min}}-1$ denote the minimum required SINR threshold of $U_n$.
The power allocation problem is formulated as
\begin{subequations}\label{eq:Pal}
\begin{align}
\text{\textbf{(P2)}}\quad
&\max_{\alpha_2}\ \mathcal F(\alpha_2)
\label{eq:Pal_a}\\
\text{\textbf{s.t.}}\quad
&C_4,\ C_5,\ C_6.
\label{eq:Pal_b}
\end{align}
\end{subequations}
The QoS requirements of $U_2$ and $U_1$ impose lower and upper bounds on $\alpha_2$, respectively. Under the assumed channel ordering, the SIC decoding condition is automatically satisfied. For numerical implementation, a small margin $\varepsilon>0$ is introduced to enforce strict power allocation inequalities. The resulting feasible interval is given by
\begin{equation}
\alpha_2^{\min}=\max\left\{\varepsilon,\frac{\beta_2}{\gamma_2}\right\},
\quad
\alpha_2^{\max}=\min\left\{\frac12-\varepsilon,
\frac{\gamma_1-\beta_1}{(1+\beta_1)\gamma_1}\right\}.
\label{eq:alpha_range}
\end{equation}
For fixed PA positions and RIS phase shifts, $\gamma_1$ and $\gamma_2$ are constants. The derivative of $\mathcal F$ with respect
to $\alpha_2$ is $\frac{\partial\mathcal F}{\partial\alpha_2}
=
\frac{(1+\gamma_1)(\gamma_2-\gamma_1)}
{(1+\alpha_2\gamma_1)^2}
\geq 0,$ where the inequality follows from the assumed channel ordering
$\gamma_2\geq\gamma_1$. Thus, $\mathcal F$ is nondecreasing in
$\alpha_2$, and the largest feasible value is optimal. Accordingly, provided that $\gamma_1>0$, $\gamma_2>0$, and $\alpha_2^{\min}\leq\alpha_2^{\max}$, the optimal power allocation coefficients for the specified numerical margin are
\begin{equation}
\alpha_2^*=\alpha_2^{\max},\quad
\alpha_1^*=1-\alpha_2^*.
\label{eq:optimal_alpha}
\end{equation}
Otherwise, the power allocation subproblem is infeasible for the fixed PA positions and RIS phase shifts.

\vspace{-0.3cm}
\subsection{RIS Phase Shift Optimization}
\label{subsec:ris_phase}
With the power allocation coefficients and PA positions fixed, we optimize the RIS phase shifts. By expressing the reflection coefficient of the $m$-th RIS element as $v_m=e^{j\theta_m}$, the unit-modulus constraint $C_3$ is automatically satisfied. The RIS phase shift optimization problem is formulated as
\begin{subequations}\label{eq:Pris}
\begin{align}
\text{\textbf{(P3)}}\quad
&\max_{\boldsymbol\theta}
\mathcal F(\boldsymbol\theta)
\label{eq:Pris_a}\\
\text{\textbf{s.t.}}\quad
&C_5,\ C_6,\ C_7.
\label{eq:Pris_b}
\end{align}
\end{subequations}
To address the non-convexity of \textbf{P3}, we employ the MM method to construct tractable surrogate problems. From \eqref{eq:effective_compact}, the $m$-th element of $\mathbf z_n=\nabla_{\boldsymbol\theta}\gamma_n$ is given by
\begin{equation}
[\mathbf z_n]_m
=-\frac{2P}{L\sigma^2}
\operatorname{Im}\left\{
g_n^*a_{n,m}e^{j\theta_m}
\right\},
\label{eq:theta_grad}
\end{equation}
where $a_{n,m}$ denotes the $m$-th element of $\mathbf a_n$. Accordingly, the objective gradient is expressed as $\nabla_{\boldsymbol\theta}\mathcal F=w_1\mathbf z_1+w_2\mathbf z_2$, where $w_1=(1-\alpha_2)(1+\alpha_2\gamma_2)/(1+\alpha_2\gamma_1)^2$ and $w_2=\alpha_2(1+\gamma_1)/(1+\alpha_2\gamma_1)$ are the partial derivatives of $\mathcal F$ with respect to $\gamma_1$ and $\gamma_2$, respectively. The QoS requirements and channel ordering constraints are represented by
\begin{equation}
\mathbf c(\boldsymbol\theta)=
\begin{bmatrix}
R_1-R_1^{\min}\\
R_2-R_2^{\min}\\
|g_2|^2-|g_1|^2
\end{bmatrix}
\succeq\mathbf 0.
\label{eq:ris_margins}
\end{equation}
At MM iteration $t$, let $\mathbf d=\boldsymbol\theta-
\boldsymbol\theta^{(t)}$ denote the phase shift update. We construct quadratic lower bounds for the objective and constraint functions as
\begin{subequations}\label{eq:mm_bounds}
\begin{align}
Q_{\mathcal F}^{(t)}(\mathbf d)
&=\mathcal F^{(t)}
+(\nabla_{\boldsymbol\theta}
\mathcal F^{(t)})^T\mathbf d
-\frac{\ell_{\mathcal F}}{2}
\|\mathbf d\|^2,
\label{eq:mm_bounds_a}\\
Q_i^{(t)}(\mathbf d)
&=c_i^{(t)}
+(\nabla_{\boldsymbol\theta}
c_i^{(t)})^T\mathbf d
-\frac{\ell_i}{2}
\|\mathbf d\|^2.
\label{eq:mm_bounds_b}
\end{align}
\end{subequations}
Here, the gradients are evaluated at $\boldsymbol\theta^{(t)}$, with
$\nabla_{\boldsymbol\theta}c_i$ obtained from $\mathbf z_1$ and
$\mathbf z_2$ using the chain rule. The positive curvature parameters
$\ell_{\mathcal F}$ and $\ell_i$ are chosen to ensure that the quadratic functions provide valid lower bounds. The resulting MM subproblem is given by
\begin{subequations}\label{eq:Pmm}
\begin{align}
\text{\textbf{(P3.1)}}\quad
&\max_{\mathbf d}
Q_{\mathcal F}^{(t)}(\mathbf d)
\label{eq:Pmm_a}\\
\text{\textbf{s.t.}}\quad
&Q_i^{(t)}(\mathbf d)\geq0,
\quad i=1,\ldots,3.
\label{eq:Pmm_b}
\end{align}
\end{subequations}
Using $\boldsymbol\theta=\boldsymbol\theta^{(t)}+\mathbf d$, problem \textbf{P3.1} can be equivalently expressed as a Euclidean projection problem. By completing the square, the unconstrained maximizer of the quadratic objective is obtained as $\mathbf z_0=\boldsymbol\theta^{(t)}+\nabla_{\boldsymbol\theta}\mathcal F^{(t)}/\ell_{\mathcal F}$. Similarly, each quadratic constraint defines a Euclidean ball
$\mathcal B_i=\{\boldsymbol\theta:
\|\boldsymbol\theta-\mathbf C_i\|^2\leq r_i^2\}$, where
\begin{equation}
\mathbf C_i=\boldsymbol\theta^{(t)}
+\tfrac{\nabla_{\boldsymbol\theta}c_i^{(t)}}{\ell_i},
\quad
r_i^2=\left\|
\tfrac{\nabla_{\boldsymbol\theta}c_i^{(t)}}{\ell_i}
\right\|^2+\tfrac{2c_i^{(t)}}{\ell_i}.
\label{eq:ris_balls}
\end{equation}
Consequently, solving \textbf{P3.1} is equivalent to projecting $\mathbf z_0$ onto $\bigcap_{i=1}^{3}\mathcal B_i$. We employ Dykstra's algorithm to compute this projection and obtain a candidate phase vector $\boldsymbol\theta^{\mathrm{cand}}$. To account for projection inaccuracies, the candidate is accepted as $\boldsymbol\theta^{(t+1)}$ only if it satisfies the original constraints and does not decrease $\mathcal F$.
Otherwise, the current phase vector is retained. The RIS reflection coefficients are updated as $v_m=e^{j\theta_m^{(t+1)}}$.

\vspace{-0.5cm}
\subsection{PA Position Optimization}
\label{subsec:pa_pos}
For fixed RIS phase shifts and power allocation coefficients, the PA position optimization problem is formulated as
\begin{subequations}\label{eq:Ppa}
\begin{align}
\text{\textbf{(P4)}}\quad
&\max_{\mathbf x^{\mathrm{pin}}}
\mathcal F(\mathbf x^{\mathrm{pin}})
\label{eq:Ppa_a}\\
\text{\textbf{s.t.}}\quad
&C_1,\ C_2,\ C_5,\ C_6,\ C_7.
\label{eq:Ppa_b}
\end{align}
\end{subequations}
To solve \textbf{P4}, we employ the sequential quadratic programming (SQP) method. Let $\mathbf x=\mathbf x^{\mathrm{pin}}$ and define $f_x(\mathbf x)=-\mathcal F(\mathbf x)$ to convert the maximization problem into a minimization problem. At SQP iteration $k$, a quadratic model of the objective and linear approximations of
the nonlinear constraints yield 
\begin{subequations}\label{eq:Psqp}
\begin{align}
\text{\textbf{(P4.1)}}\quad
&\min_{\mathbf d_x}\;
(\nabla_{\mathbf x}f_x^{(k)})^T\mathbf d_x
+\frac12\mathbf d_x^T
\mathbf B^{(k)}\mathbf d_x
\label{eq:Psqp_a}\\
\text{\textbf{s.t.}}\quad
&\mathbf c_x^{(k)}
+(\mathbf J_x^{(k)})^T\mathbf d_x
\preceq\mathbf0,
\label{eq:Psqp_b}\\
&\mathbf A(\mathbf x^{(k)}+\mathbf d_x)
+\Delta\mathbf1\preceq\mathbf0,
\label{eq:Psqp_c}\\
&\mathbf0\preceq\mathbf x^{(k)}+\mathbf d_x
\preceq D_1\mathbf1.
\label{eq:Psqp_d}
\end{align}
\end{subequations}
Here, $\mathbf d_x$ denotes the PA position update, and $\mathbf B^{(k)}$ is a positive definite approximation of the Hessian of
the Lagrangian. The nonlinear constraint vector is defined as $\mathbf c_x=-\mathbf c$, where $\mathbf c$ contains the three constraint margins in \eqref{eq:ris_margins}, evaluated with respect to $\mathbf x$. The matrix $\mathbf J_x\in\mathbb R^{L\times3}$ contains the gradients of the components of $\mathbf c_x$ with respect to $\mathbf x$. 
The matrix $\mathbf A\in\mathbb R^{(L-1)\times L}$ encodes the adjacent PA spacing constraints, with consecutive entries $1$ and $-1$ in each row and zeros elsewhere. The remaining constraints restrict the PA positions to the waveguide. Since $\mathbf B^{(k)}$ is positive definite and all constraints are affine, \textbf{P4.1} is a convex quadratic
programming (QP) problem. To compute the objective gradient, the effective channel in \eqref{eq:effective_compact} is rewritten as $g_n=\sum_{l=1}^{L}e^{-j\varphi_l}H_{l,n}$, where $H_{l,n}$ comprises the direct and RIS-reflected channel components associated with the $l$-th PA, excluding the in-waveguide phase shift. With the waveguide feed at $(0,0,h)$, the channel derivative is given by
\begin{equation}
\frac{\partial g_n}
{\partial x_l^{\mathrm{pin}}}
=e^{-j\varphi_l}
\left(
\frac{\partial H_{l,n}}
{\partial x_l^{\mathrm{pin}}}
-j\frac{2\pi}{\lambda_g}H_{l,n}
\right).
\label{eq:pa_channel_grad}
\end{equation}
Accordingly, the $l$-th element of $\nabla_{\mathbf x}\gamma_n$ is obtained as $\partial\gamma_n/\partial x_l^{\mathrm{pin}}=(2P/(L\sigma^2))\operatorname{Re}\{g_n^*\partial g_n/\partial x_l^{\mathrm{pin}}\}$. Using the chain rule, the objective gradient is given by $\nabla_{\mathbf x}f_x=-w_1\nabla_{\mathbf x}\gamma_1-w_2\nabla_{\mathbf x}\gamma_2$, where $w_1$ and $w_2$ are defined in Subsection~\ref{subsec:ris_phase}. The constraint gradients are similarly obtained from the SNR gradients. At each SQP iteration, the convex QP subproblem \textbf{P4.1} is solved using MATLAB's \texttt{fmincon} function with the SQP algorithm to obtain $\mathbf d_x$. The PA positions are then updated as $\mathbf x^{(k+1)}=\mathbf x^{(k)}+\mathbf d_x$. The resulting SQP procedure seeks a stationary solution of \textbf{P4}.

The three optimization stages are combined in the AO algorithm summarized in Algorithm~\ref{alg:AO}. Let $\boldsymbol\xi=(\alpha_2,\mathbf x^{\mathrm{pin}},\mathbf v)$ denote the optimization variables and $\mathcal S_{\mathrm{P1}}$ the feasible set of \textbf{P1}. Starting from a feasible solution satisfying the numerical power allocation margin, the algorithm accepts only feasible updates that do not decrease $\mathcal F$. The resulting sequence of objective values is nondecreasing and bounded above, and hence converges.
Regarding computational complexity, the closed-form power allocation in \eqref{eq:optimal_alpha} requires $\mathcal O(1)$ operations given the effective channel gains. The MM-based RIS phase shift optimization requires $\mathcal O(T_{\rm MM}M)$ operations assuming precomputed channel coefficients and fixed numbers of Dykstra cycles and $\mathcal O(M)$ curvature-selection trials. The SQP-based PA position optimization requires approximately $\mathcal O(T_{\rm SQP}(L^3+LM))$ operations, accounting for dense QP solutions with bounded solver iterations and channel and gradient evaluations. Here, $T_{\rm MM}$ and $T_{\rm SQP}$ denote the respective iteration counts. Including the $\mathcal O(LM)$ channel preprocessing cost, the overall complexity for $I_{\rm AO}$ AO iterations is estimated as
$
\mathcal O\!\left(
I_{\rm AO}\left[
T_{\rm MM}M+
T_{\rm SQP}(L^3+LM)
\right]\right).
$

\begin{algorithm}[t]
\caption{Proposed Algorithm for Solving \textbf{P1}}
\label{alg:AO}
\begin{algorithmic}[1]
\STATE \textbf{Initialize:} Set $i=0$,
a feasible solution $\boldsymbol\xi^{(0)}$,
convergence tolerance $\epsilon_{\rm AO}$,
and maximum iteration number $I_{\max}$.
\STATE Calculate $\mathcal F^{(0)}$.
\REPEAT
\STATE $i=i+1$.
\STATE Update $\alpha_2^{(i)}$ using
\eqref{eq:optimal_alpha}.
\STATE Update $\mathbf v^{(i)}$ by applying
the MM method to \textbf{P3}.
\STATE Update $\mathbf x^{\mathrm{pin},(i)}$
by applying the SQP method to \textbf{P4}.
\STATE Calculate $\mathcal F^{(i)}$.
\IF{$\boldsymbol\xi^{(i)}
\in\mathcal S_{\mathrm{P1}}$
and $\mathcal F^{(i)}\geq\mathcal F^{(i-1)}$}
\STATE $\boldsymbol\xi
\leftarrow\boldsymbol\xi^{(i)}$.
\ELSE
\STATE $\boldsymbol\xi
\leftarrow\boldsymbol\xi^{(i-1)}$.
\STATE \textbf{break}.
\ENDIF
\UNTIL{$\mathcal F^{(i)}-\mathcal F^{(i-1)}
\leq\epsilon_{\rm AO}$ or $i\geq I_{\max}$.}
\STATE Obtain $\alpha_2^*$ using
\eqref{eq:optimal_alpha}
for the retained $\mathbf v$
and $\mathbf x^{\mathrm{pin}}$.
\STATE \textbf{Output:}
$\alpha_2^*$, $\mathbf v$,
$\mathbf x^{\mathrm{pin}}$.
\end{algorithmic}
\end{algorithm}

\vspace{-0.4cm}
\section{Numerical Results}


\begin{figure*}[!t]
    \centering
    \captionsetup[subfigure]{skip=2pt}

    \begin{subfigure}[b]{0.247\textwidth}
        \centering
        \makebox[\linewidth][c]{%
        \includegraphics[width=1.03\linewidth,height=3.70cm,keepaspectratio]{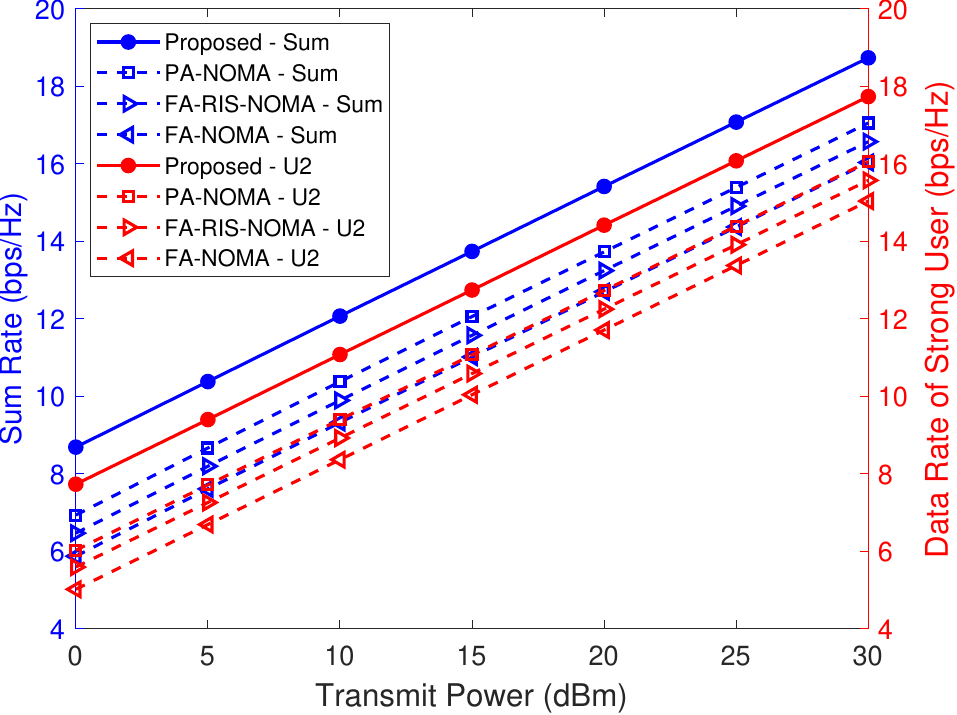}}
        \caption{}
        \label{fig:power_comparison}
    \end{subfigure}%
    \begin{subfigure}[b]{0.247\textwidth}
        \centering
        \makebox[\linewidth][c]{%
        \includegraphics[width=1.00\linewidth,height=3.50cm,keepaspectratio]{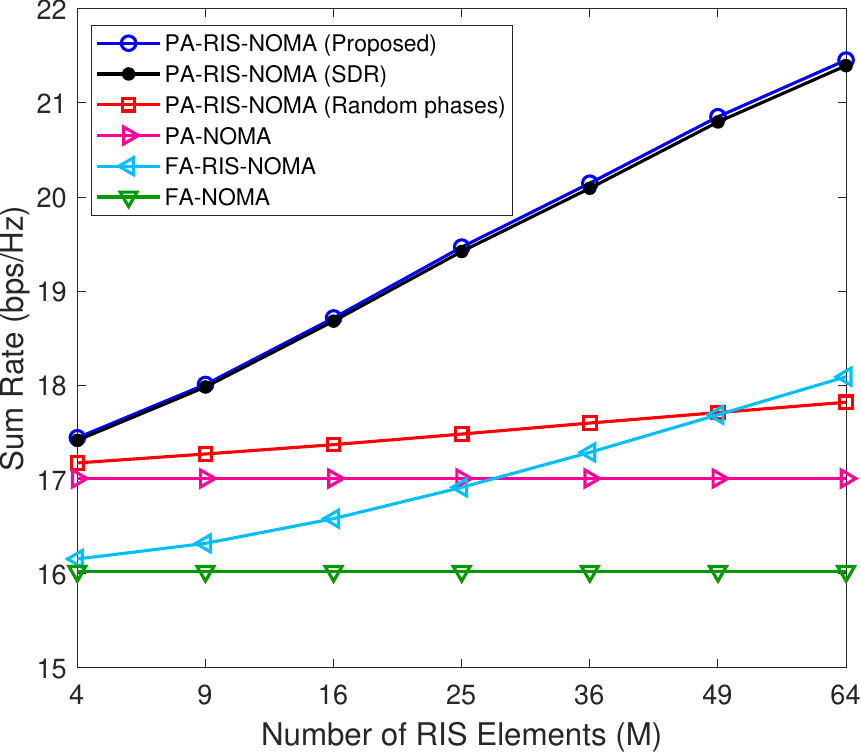}}
        \caption{}
        \label{fig:ris_elements}
    \end{subfigure}%
    \begin{subfigure}[b]{0.247\textwidth}
        \centering
        \makebox[\linewidth][c]{%
        \includegraphics[width=1.00\linewidth,height=3.50cm,keepaspectratio]{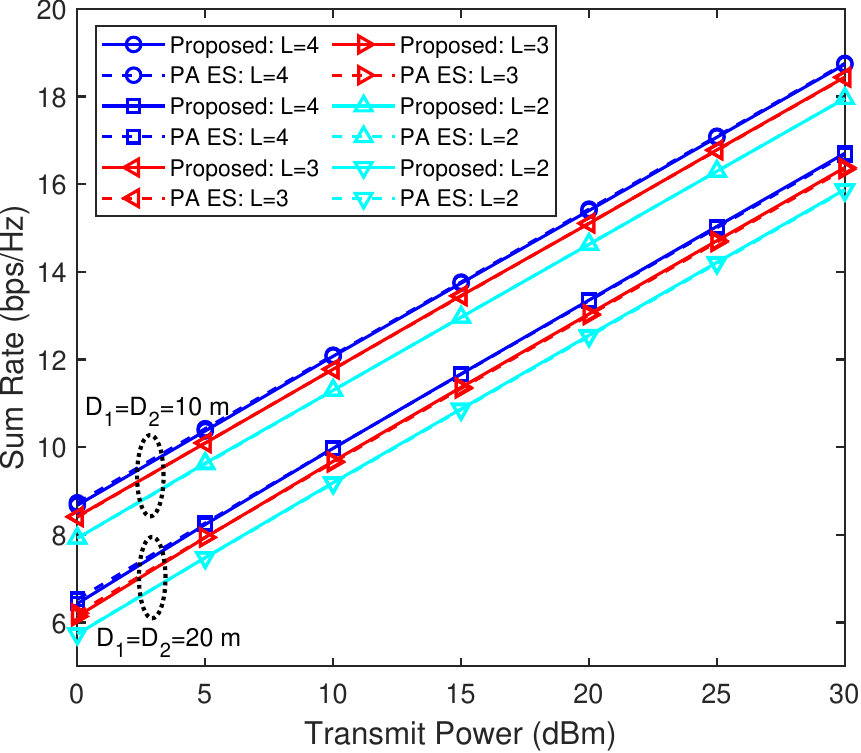}}
        \caption{}
        \label{fig:pa_es}
    \end{subfigure}%
    \begin{subfigure}[b]{0.247\textwidth}
        \centering
        \makebox[\linewidth][c]{%
        \includegraphics[width=1.00\linewidth,height=3.50cm,keepaspectratio]{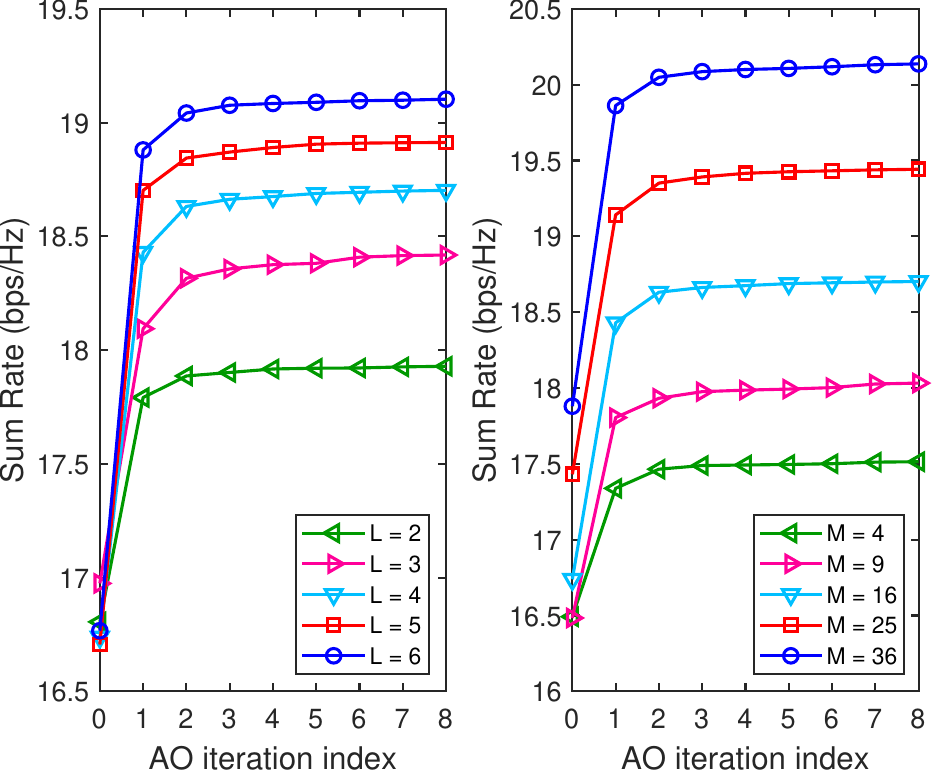}}
        \caption{}
        \label{fig:ao_convergence}
    \end{subfigure}

    \vspace{-4pt}
    \caption{Performance evaluation of the proposed RIS-assisted PA-NOMA system (a) sum rate and $U_2$ rate versus transmit power ($L=4$, $M=16$) (b) sum rate versus the number of RIS elements ($L=4$, $P=30$ dBm) (c) sum rate versus transmit power for $D_1=D_2\in\{10,20\}$ m ($M=16$) and (d) AO convergence for different numbers of PAs ($L$) and RIS elements ($M$).}
    \label{fig:results}
    \vspace{-10pt}
\end{figure*}

This section evaluates the achievable sum-rate performance and convergence behavior of the proposed PA-RIS-NOMA algorithm against several benchmark schemes. Following the simulation parameters in~\cite{Xu2025DownlinkPASS,Fu2026RISPASS,Shen2021RISSpacing}, the noise power is set to $-90$~dBm, the carrier frequency to $f_c=28$~GHz, the minimum PA and RIS element spacings to $\Delta=d=\lambda/2$, the effective refractive index to $n_{\mathrm{eff}}=1.4$ and the waveguide height to $h=3$~m. The users are uniformly distributed within a $D_1\times D_2=10\times10$~m$^2$ region, and the RIS reference element is located at $\boldsymbol\psi^{\mathrm{ris}}=(D_1,D_2,h)$. The minimum rate requirement is set to $0.5$~bps/Hz for each user.

Fig.~\ref{fig:power_comparison} compares the proposed algorithm with three benchmarks. PA-NOMA operates without an RIS, with PA positions and power allocation optimized as in~\cite{Zhou2025NOMAPASS}. FA-NOMA employs an FA transmitter at $(0,0,h)$ with maximum-ratio transmission (MRT) beamforming~\cite{Wu2019RISBeamforming} and optimized NOMA power allocation. FA-RIS-NOMA incorporates an RIS into this configuration, with phase shifts optimized using MM. As transmit power increases, both the sum rate and strong-user rate improve across all schemes, with the proposed method consistently achieving the highest values. The gains over PA-NOMA and FA-RIS-NOMA demonstrate the benefits of RIS assistance and flexible PA positioning. Joint optimization of PA positions and RIS phases enables more favorable combining of direct and RIS-reflected signals at the receivers. Moreover, the strong-user rate accounts for most of the sum rate, suggesting that the sum-rate gains are primarily reflected in strong-user throughput under the QoS constraints.

Fig.~\ref{fig:ris_elements} examines the effect of the number of RIS elements $M$. For comparison, the MM-based RIS phase optimization is replaced by SDR~\cite{Li2022RISNOMADirect}, implemented using the MATLAB \texttt{CVX} toolbox, while the other optimization stages are unchanged. The MM-based method achieves nearly the same sum-rate performance as SDR, with both methods outperforming the random-phase configuration and other benchmarks. The performance gap between optimized and random RIS phases widens as $M$ increases, reflecting the additional phase-control degrees of freedom and improved constructive combining of reflected signals. Although FA-RIS-NOMA benefits from increasing $M$, it remains below the proposed architecture, highlighting the advantage of jointly optimizing PA positions and RIS phases. In contrast, PA-NOMA and FA-NOMA remain unaffected by $M$, as neither employs an RIS.

Fig.~\ref{fig:pa_es} evaluates the SQP-based PA position optimization against grid-based exhaustive search (PA ES) for $L=2,3,4$ and two deployment areas. In PA ES, the SQP stage is replaced by exhaustive search over feasible grid positions, while the remaining optimization stages are unchanged. The proposed method achieves sum rates close to those of PA ES across the configurations, indicating that SQP can identify effective PA positions without evaluating all grid configurations. Increasing the number of PAs improves the sum rate by providing greater flexibility in shaping the effective channels. Conversely, expanding the deployment area reduces achievable rates due to greater propagation losses, which cannot be fully compensated by PA positioning and RIS phase optimization.

Fig.~\ref{fig:ao_convergence} presents the convergence behavior of the proposed AO algorithm for varying numbers of PAs (left, $M=16$) and RIS elements (right, $L=4$). In both cases, the sum rate increases monotonically, with the largest improvements occurring during the initial iterations, followed by progressively smaller gains. Increasing either $L$ or $M$ leads to higher achievable sum rates. Despite the different system configurations, all curves stabilize within approximately eight AO iterations, indicating rapid convergence in terms of iteration count.

\vspace{-0.2cm}
\section{Conclusion}
This paper investigates QoS-constrained sum-rate maximization for a two-user RIS-assisted downlink PASS-NOMA system. To address the joint optimization of PA positions, RIS phase shifts, and power allocation coefficients, an AO algorithm  is developed using SQP, MM, and a closed-form solution, respectively. Numerical results  demonstrate that the proposed approach outperforms the considered benchmarks, while the MM and SQP stages achieve performance comparable to SDR and grid-based PA ES, respectively. 

\vspace{-0.2cm}
\bibliographystyle{IEEEtran}
\bibliography{references1}

\end{document}